\documentclass[aps,prd,showpacs,nofootinbib,floats,floatfix,preprintnumbers,onecolumn]{revtex4}

\usepackage{bm}
\usepackage{latexsym}
\usepackage{dcolumn}
\usepackage{amsfonts,amssymb,amsmath}
\usepackage{graphicx,epsfig}
\usepackage{psfrag}
\usepackage{multirow}

\def\beq{\begin{equation}}
\def\eeq{\end{equation}}
\def\bea{\begin{eqnarray}}
\def\eea{\end{eqnarray}}
\def\benu{\begin{enumerate}}
\def\eenu{\end{enumerate}}
\def\nn{\nonumber}
\def\l{\left}
\def\r{\right}
\def\d{{\rm d}}

\def\pa{\partial}
\def\f{\frac}

\begin{document}

\title{A note on perfect scalar fields}
\author{Sanil Unnikrishnan$^1$}\email[E-mail:~]{sanil@iucaa.ernet.in}
\author{L.~Sriramkumar$^2$}\email[E-mail:~]{sriram@hri.res.in}
\affiliation{$^{1}$IUCAA, Post Bag 4, Ganeshkhind, Pune 411 007,
India.\\
$^{2}$Harish-Chandra Research Institute, Chhatnag Road, Jhunsi,
Allahabad~211~019, India.}
\date{\today}
\begin{abstract}
We derive a condition on the Lagrangian density describing a generic,
single, non-canonical scalar field, by demanding that the intrinsic,
non-adiabatic pressure perturbation associated with the scalar field
vanishes identically.
Based on the analogy with perfect fluids, we refer to such fields as
perfect scalar fields.
It is common knowledge that models that depend only on the kinetic energy
of the scalar field (often referred to as pure kinetic models) possess
no non-adiabatic pressure perturbation.
While we are able to construct models that seemingly depend on the
scalar field and also do not contain any non-adiabatic pressure
perturbation, we find that all such models that we construct allow a
redefinition of the field under which they reduce to pure kinetic models.
We show that, if a perfect scalar field drives inflation, then, in such
situations, the first slow roll parameter will always be a monotonically
decreasing function of time.
We point out that this behavior implies that these scalar fields can not
lead to features in the inflationary, scalar perturbation spectrum.
\end{abstract}
\maketitle


\section{Introduction}

In cosmology, the sources of matter that drive the expansion of the
universe are often considered to be either fluids or, in particular,
while considering the inflationary epoch or late time acceleration,
as scalar fields.
Almost always, the fluids are assumed to be perfect, i.e. it is assumed
that they do not possess any intrinsic, non-adiabatic pressure perturbation.
In contrast, generically, the non-adiabatic pressure perturbation proves
to be non-zero for scalar fields.
It is well known that iso-curvature (i.e. the extrinsic, non-adiabatic
pressure) perturbations are always present when one considers multiple
fields and/or fluids.
In the context of inflation, it is often said that single scalar field
models are adiabatic (see any of the following texts~\cite{texts} or
reviews~\cite{reviews}).
While this is true, it is not  because the non-adiabatic pressure
perturbation is identically zero for the single field models, but
because they decay exponentially for cosmological modes after they
leave the Hubble radius during inflation (for a discussion on
specific cases, see Refs.~\cite{bep-shs}; for a generic discussion,
see Ref.~\cite{christopherson-2009}).
It is interesting to enquire whether there exist scalar field models
wherein the intrinsic, non-adiabatic pressure perturbation vanishes
exactly (say, at the first order in the perturbations), as in the case
of perfect fluids.

In this paper, using the standard definition of the non-adiabatic
pressure perturbation associated with the scalar fields, we shall
obtain a condition on the Lagrangian density describing the scalar
field by demanding that the non-adiabatic pressure perturbation is
identically zero at the first order in the perturbation theory.
We shall also discuss specific examples of scalar field models that
satisfy this condition.
Following the convention with fluids, we shall call these fields as
perfect scalar fields.
It is well known that models that depend only on the kinetic energy
of the scalar field behave as perfect fluids.
We are also able to construct models which seem to depend on the scalar
field, and possess no non-adiabatic pressure perturbation.
However, we find that all such models that we consider reduce to pure
kinetic models after a suitable redefinition of the field.
Interestingly, we shall show that, if the perfect scalar fields are used
to drive inflation, they can not lead to features in the inflationary,
scalar perturbation spectrum.

This paper is organized as follows.
In the following section, we shall rapidly outline essential,
{\it linear},\/ cosmological perturbation theory.
We shall highlight the key equations, and point out the standard
definition of the intrinsic, non-adiabatic pressure perturbation.
In Sec.~\ref{sec:condition}, by demanding that the non-adiabatic
pressure perturbation associated with the scalar field vanishes
exactly, we shall obtain a condition on the Lagrangian density of
the scalar field.
We shall discuss a few specific examples of Lagrangian densities
that satisfy this condition in Sec.~\ref{sec:examples}.
In Sec.~\ref{sec:ii}, we shall show that, if scalar fields with a
vanishing non-adiabatic pressure perturbation are responsible for
inflation then, in such cases, the first slow roll parameter is
always a monotonically decreasing function of time.
We shall argue that this behavior points to the fact that such
inflatons will not lead to features in the primordial scalar
perturbation spectrum.
Finally, we shall close with a few concluding remarks in
Sec.~\ref{sec:cr}.

A few words on our conventions and notations are in order before we
proceed.
We shall set $c=1$, but shall display $G$ explicitly.
We shall work in a spatially flat, $(3 + 1)$-dimensional Friedmann
background, and we shall adopt the metric signature of $(+,-,-,-)$.
Note that, while the Greek indices $\mu$ or $\nu$ shall denote the
spacetime coordinates, the Latin indices $i$ or $j$ shall denote
the spatial coordinates.
The sub-script $k$ shall refer to the Fourier component of the
perturbations.
We shall express all the quantities in terms of either the cosmic or
the conformal time coordinates.
While an overdot shall denote differentiation with respect to the
cosmic time, an overprime shall denote differentiation with respect
to the conformal time.
It is handy to note that, for any given function, say, $f$,
${\dot f}=(f'/a)$ and ${\ddot f}=\l[\l(f''/a^2\r)-\l(f'\,
a'/a^{3}\r)\r]$, where $a$ is the scale factor associated with the
Friedmann metric.


\section{Scalar perturbations in a Friedmann universe---Key equations
and definitions}\label{sec:cpt}

In this section, using the equations governing the evolution of the
scalar perturbations induced by an arbitrary matter field and the
standard definition of the non-adiabatic pressure perturbation, we
shall arrive at an expression for the non-adiabatic pressure
perturbation associated with the scalar field in terms of the
adiabatic and the effective speeds of sound associated with the
perturbations.


In $(3 + 1)$-dimensions, when no perturbations are present, the spatially
flat Friedmann universe is described by the line element
\beq
\d s^2 = \d t^2-a^{2}(t)\; \d{\bf x}^2
= a^{2}(\eta)\, \l(\d\eta^{2} - \d{\bf x}^2\r),\label{eq:f-le}
\eeq
where $t$ is the cosmic time, $a(t)$ is the scale factor, and $\eta=\int
[\d t/a(t)]$ denotes the conformal time.
If $\rho$ and $p$ denote the energy density and the pressure of the smooth
component of the matter field that is driving the expansion, then the
Einstein's equations for the above line-element lead to the following
Friedmann equations for the scale factor~$a(t)$:
\beq
H^2= \l(\frac{8\,\pi\, G}{3}\r)\, {\rho}\qquad{\rm and}\qquad
\l(\frac{\ddot a}{a}\r)
= -\l(\frac{4\,\pi\, G}{3}\r)\, \l({\rho}+3\, p\r),
\eeq
where $H=({\dot a}/a)$ is the Hubble parameter.


\subsection{Equations governing the scalar perturbations and the definition
of the non-adiabatic pressure perturbation}

If we now take into account the scalar perturbations to the background
metric~(\ref{eq:f-le}), then the Friedmann line-element, in general,
can be written as~\cite{texts,reviews}
\beq
\d s^2
= (1+2\, A)\,\d t ^2 - 2\, a(t)\, (\pa_{i} B )\; \d t\; \d x^i\,
-a^{2}(t)\; \l[(1-2\, \psi)\; \delta _{ij}+ 2\, \l(\pa_{i}\, \pa_{j}E \r)\,
\d x^i\, \d x^j\r],\label{eq:f-le-sp}
\eeq
where $A$, $B$, $\psi$ and $E$ are the scalar functions that describe
the perturbations.
The gauge-invariant Bardeen variables that characterize the
two degrees of freedom describing the scalar perturbations are
given by~\cite{bardeen-1980}
\beq
\Phi \equiv A + \l[a\, (B-a\, {\dot E})\r]^{\cdot}
\qquad {\rm and}\qquad
\Psi
\equiv \psi - \l[a\, H\, (B-a\, {\dot E})\r].\label{eq:bps}
\eeq
In the absence of anisotropic stresses, as it is in the case of
the scalar field sources that we shall be interested in, it can
be readily shown that, at the linear order in the perturbations,
the non-diagonal, spatial components of the Einstein's equations
lead to the relation: $\Phi=\Psi$.
The remaining first order Einstein's equations then reduce
to~\cite{texts,reviews}
\bea
\l(\f{1}{a^{2}}\r)\, \nabla^2 \Phi
-3\, H\, \l({\dot \Phi} +H\, \Phi\r)
&=& \l(4\, \pi\, G\r)\, \l[\delta \rho+ ({\dot \rho}\, a)\;
(B-a\, {\dot E})\r]
= \l(4\, \pi\, G\r)\, \widehat{\delta \rho},\label{eq:fo-ee-00}\\
\pa_{i}\l({\dot \Phi}+ H\, \Phi\r)
&=& \l(4\,\pi\, G\r)\,
\pa_{i}\l[\delta \sigma+ [(\rho+p)\, a]\, (B-a\, {\dot E})\r]
= \l(4\, \pi\, G\r)\,
\l(\pa_{i}\;\widehat{\delta \sigma}\r),\;\;\label{eq:fo-ee-0i}\\
{\ddot \Phi}+ 4\, H\, {\dot \Phi}
+\l(2\, {\dot H}+ 3\, H^2\r)\,\Phi\;
&=& \l(4\, \pi\, G\r)\,
\l[\delta p+ ({\dot p}\, a)\; (B-a\, {\dot E})\r]
=\l(4\, \pi\, G\r)\, \widehat{\delta p},\label{eq:fo-ee-ii}
\eea
where $\delta\rho$, $\delta \sigma$ and $\delta p$ denote the
perturbations at the linear order in the energy density, flux,
and the pressure of the matter field, respectively, while the
quantities $\widehat{\delta \rho}$, $\widehat{\delta \sigma}$
and $\widehat{\delta p}$ represent the corresponding
gauge-invariant quantities.
The first and the third of the above first order Einstein equations
can be combined to lead to the following differential equation for the
Bardeen potential $\Phi$~\cite{texts,reviews}:
\beq
\Phi^{\prime\prime}
+3\, {\mathcal H}\, \l(1+c_{_{\rm A}}^2\r)\, \Phi^{\prime}
-c_{_{\rm A}}^2\, \nabla^{2}\Phi
+ \l[2\, {\mathcal H}'+ \l(1+3\, c_{_{\rm A}}^2\r)\,
{\mathcal H}^2\r]\, \Phi
= \l(4\, \pi\, G\,\, a^2\r)\, \delta p^{_{\rm NA}},\label{eq:em-bp}
\eeq
where ${\cal H}=(H\, a)$ is the conformal Hubble parameter.
In arriving at this equation, we have changed over to the conformal
time coordinate, and have made use of the following standard definition
of the non-adiabatic pressure
perturbation~$\delta p^{_{\rm NA}}$~\cite{gordon-2001}:
\beq
\delta p^{_{\rm NA}}=\l(\widehat{\delta p}-
c_{_{\rm A}}^2\, \widehat{\delta\rho}\r)
= \l(\delta p- c_{_{\rm A}}^2\, \delta\rho\r),\label{eq:delta-p-na-gen}
\eeq
where $c_{_{\rm A}}\equiv\sqrt{\l(p'/\rho'\r)}$ denotes the adiabatic
speed of the perturbations.


\subsection{Perturbations induced by a generic scalar field}

Consider a generic scalar field~$\phi$ that is described by the
action~\cite{ki}
\beq
S[\phi]=\int\!\d^{4}x\, \sqrt{-g}\; {\cal L}(X,\phi),
\eeq
where $X$ is a term that describes the kinetic energy of the scalar
field, and is defined as
\beq
X=\l(\f{1}{2}\r)\, \l(\pa_{\mu}\phi\; \pa^{\mu}\phi\r).
\eeq
Let us assume that the Lagrangian density ${\cal L}$ is an arbitrary
function of the kinetic term $X$ and the field $\phi$.
The stress-energy tensor associated with the above action can be written
as
\beq
T^{\mu}_{\nu}
=\l(\f{\pa{\cal L}}{\pa X}\r)\, \l(\pa^{\mu}\phi\; \pa_{\nu}\phi\r)
- \delta^{\mu}_{\nu}\, {\cal L}.\label{eq:set}
\eeq

When no perturbations are present, the energy density and the pressure
associated with the homogeneous scalar field are given by
\beq
\rho = \l(\f{\pa {\cal L}}{\pa X}\r)\, (2\, X)- {\cal L}
\qquad{\rm and}\qquad
p = {\cal L},\label{eq:rho-p-gsf}
\eeq
with $X=({\dot \phi}^{2}/2)$.
If we now denote the perturbation in the scalar field as $\delta \phi$,
then, the perturbations in the energy density, the momentum flux and
the pressure of the scalar field can be obtained to be~\cite{ki,sanil-2008}
\bea
\delta\rho
&=& \l[\l(\f{\pa {\cal L}}{\pa X}\r)
+ \l(2\, X\r)\, \l(\f{\pa^{2} {\cal L}}{\pa X^{2}}\r)\r]\,
{\dot \phi}\,\l(\dot{\delta\phi} - {\dot \phi}\,A\r)
- \l[\l(\f{\pa {\cal L}}{\pa \phi}\r)
- \l(2\, X\r)\,
\l(\f{\pa^{2} {\cal L}}{\pa X\, \pa\phi}\r)\r]\,\delta\phi,
\label{eq:delta-rho-gsf}\\
\delta\sigma
&=& \l(\f{\pa {\cal L}}{\pa X}\r)\,
\l({\dot \phi}\; \delta \phi\r),
\label{eq:delta-sigma-gsf}\\
\delta p
&=& \l(\f{\pa {\cal L}}{\pa X}\r)\,
{\dot \phi}\,\l(\dot{\delta\phi} - {\dot \phi}\, A\r)
+\l(\f{\pa {\cal L}}{\pa \phi}\r)\, \delta\phi.
\label{eq:delta-p-gsf}
\eea
The gauge-invariant perturbation in the scalar field, say, $\delta
\varphi$, is given by~\cite{texts,reviews}
\beq
\delta \varphi
= \l[\delta \phi+ ({\dot \phi}\; a)\; (B-a\, {\dot E})\r].
\eeq
The gauge-invariant perturbations in the energy density, the momentum
flux and the pressure of the scalar field can be expressed in terms of
the gauge-invariant perturbation in the scalar field~$\delta \varphi$
and the Bardeen potential~$\Phi$ as follows:
\bea
\widehat{\delta\rho}
&=&  \l[\l(\f{\pa {\cal L}}{\pa X}\r)
+ \l(2\, X\r)\,\l(\f{\pa^{2} {\cal L}}{\pa X^{2}}\r)\r]\,
{\dot \phi}\,\l(\dot{\delta{\varphi}}
- {\dot \phi}\, \Phi\r)\,
- \l[\l(\f{\pa {\cal L}}{\pa \phi}\r)
- \l(2\, X\r)\,\l(\f{\pa^{2} {\cal L}}{\pa X\, \pa\phi}\r)\r]\,
\delta{\varphi},
\label{eq:gi-delta-rho-gsf}\\
\widehat{\delta\sigma}
&=& \l(\f{\pa {\cal L}}{\pa X}\r)\,
\l({\dot \phi}\; \delta \varphi\r),
\label{eq:gi-delta-sigma-gsf}\\
\widehat{\delta p}
&=& \l(\f{\pa {\cal L}}{\pa X}\r)\,
{\dot \phi}\,\l(\dot{\delta\varphi} - {\dot \phi}\,\Phi\r)
+\l(\f{\pa {\cal L}}{\pa \phi}\r)\, \delta\varphi.
\label{eq:gi-delta-p-gsf}
\eea

Our goal now is to arrive at an expression for the non-adiabatic
pressure perturbation $\delta p^{_{\rm NA}}$ for the scalar field
in terms of the Bardeen potential~$\Phi$.
Algebraically, we find that, one useful way would be to derive the
equivalent of the Bardeen equation~(\ref{eq:em-bp}) for the case of
the scalar field.
On substituting the above expressions for~$\widehat{\delta \rho}$
and $\widehat{\delta p}$ in the first order Einstein's
equations~(\ref{eq:fo-ee-00}) and ({\ref{eq:fo-ee-ii}), we find
that the Bardeen potential $\Phi$ induced by the perturbations in the
scalar field satisfies the following differential equation:
\beq
\Phi^{\prime\prime}
+3\, {\mathcal H}\, \l(1+c_{_{\rm A}}^2\r)\, \Phi^{\prime}
-c_{_{\rm A}}^2\, \nabla^{2}\Phi
+ \l[2\, {\mathcal H}'+ \l(1+3\, c_{_{\rm A}}^2\r)\,
{\mathcal H}^2\r]\, \Phi= \l(c_{_{\rm S}}^2-c_{_{\rm A}}^2\r)\,
\nabla^{2}\Phi,\label{eq:em-bp-gsf}
\eeq
where the quantity $c_{_{\rm S}}$ is often referred to as the
effective speed of the perturbations, and is given by~\cite{ki}
\beq
c_{_{\rm S}}^{2}
= \l[\f{\l({\pa {\cal L}}/{\pa X}\r)}{\l({\pa {\cal L}}/{\pa X}\r)
+ \l(2\, X\r)\, \l({\pa^{2} {\cal L}}/{\pa X^{2}}\r)}\r].
\label{eq:cs2}
\eeq
It should be mentioned that, in arriving at the above equation for the
Bardeen potential~$\Phi$, we have made use of the following equation of
motion for the background scalar field~$\phi$:
\beq
\l[\l(\f{\pa {\cal L}}{\pa X}\r)
+ \l(2\, X\r)\, \l(\f{\pa^{2}{\cal L}}{\pa X^{2}}\r)\r]\,
{\ddot \phi}+\l[\l(3\, H\r)\, \l(\f{\pa {\cal L}}{\pa X}\r)
+ {\dot \phi}\, \l(\f{\pa^{2}{\cal L}}{\pa X\, \pa\phi}\r)\r]\,
{\dot \phi}
-\l(\f{\pa {\cal L}}{\pa \phi}\r)=0,\label{eq:em-gsf}
\eeq
which, in turn, can be derived from the equation describing the
conservation of the background energy density, viz.
\beq
{\dot \rho}+\l(3\, H\r)\; \l(\rho+p\r)=0,\label{eq:ce}
\eeq
and using the expressions for~$\rho$ and~$p$ from Eq.~(\ref{eq:rho-p-gsf}).
Upon comparing the equations~(\ref{eq:em-bp}) and~(\ref{eq:em-bp-gsf}),
it is straightforward to see that the non-adiabatic pressure perturbation
$\delta p^{_{\rm NA}}$ for the scalar field can be expressed as
\beq
\delta p^{_{\rm NA}}
= \l(\f{c_{_{\rm S}}^{2} - c_{_{\rm A}}^{2}}{4\, \pi\, G\, a^{2}}\r)\;
\nabla^{2}\Phi.\label{eq:delta-p-na-gsf}
\eeq

This result for $\delta p^{_{\rm NA}}$ can be obtained in a more
straightforward manner by first noticing that the
quantities~$\widehat{\delta \rho}$, $\widehat{\delta \sigma}$
and~$\widehat{\delta p}$ as given by the
expressions~(\ref{eq:gi-delta-rho-gsf}), (\ref{eq:gi-delta-sigma-gsf})
and~(\ref{eq:gi-delta-p-gsf}), respectively, are related as follows:
\beq
\widehat{\delta p}
=\l[c_{_{\rm S}}^{2}\; \widehat{\delta \rho}
+\l(3\, H\r)\; \l(c_{_{\rm S}}^{2}-c_{_{\rm A}}^{2}\r)\;
\widehat{\delta \sigma}\r].
\eeq
This relation, in turn, allows us to write $\delta p^{_{\rm NA}}$ as
\beq
\delta p^{_{\rm NA}}
=\l(\widehat{\delta p}-
c_{_{\rm A}}^2\, \widehat{\delta\rho}\r)
= \l(c_{_{\rm S}}^{2} - c_{_{\rm A}}^{2}\r)\,
\l[\widehat{\delta \rho} + \l(3\, H\r)\; \widehat{\delta \sigma}\r].
\eeq
Also, the first order Einstein's equations~(\ref{eq:fo-ee-00})
and~({\ref{eq:fo-ee-0i}) can be combined to arrive at
\beq
\l[\widehat{\delta \rho} + \l(3\, H\r)\; \widehat{\delta \sigma}\r]
=\l(\f{1}{4\, \pi\, G\, a^{2}}\r)\;
\nabla^{2}\Phi.\label{eq:c12}
\eeq
Evidently, these last two equations immediately lead to the
expression~(\ref{eq:delta-p-na-gsf}) for $\delta p^{_{\rm NA}}$.

The reason for the appearance of the quantity $c_{_{\rm S}}^2$ in this
expression for $\delta p^{_{\rm NA}}$ can be understood as follows.
Let us work in a gauge wherein $\delta \phi=0$, a choice of coordinates
that is often referred to as the comoving gauge~\cite{reviews}.
In such a gauge, no flux of energy arises, i.e. $^{_{\rm (C)}}\delta
\sigma=0$ [cf. Eq.~(\ref{eq:delta-sigma-gsf})], and we have used the
super-script {\scriptsize ${\rm (C)}$} to denote the fact that we are
working in the comoving gauge.
It is clear from the expressions~(\ref{eq:delta-rho-gsf})
and~(\ref{eq:delta-p-gsf}) for $\delta \rho$ and $\delta p$ that, in the
comoving gauge, we have
\beq
^{_{\rm (C)}}\delta p
= c_{_{\rm S}}^{2}\; ^{_{\rm (C)}}\delta\rho.
\eeq
Therefore, the non-adiabatic pressure perturbation $\delta p^{_{\rm NA}}$
in this gauge is given by
\beq
\delta p^{_{\rm NA}}
= \l[{}^{_{\rm (C)}}\delta p
- c_{_{\rm A}}^{2}\; ^{_{\rm (C)}}\delta\rho\r]
=\l(c_{_{\rm S}}^{2} - c_{_{\rm A}}^{2}\r)\;\,
^{_{\rm (C)}}\delta\rho
\eeq
and, since $^{_{\rm (C)}}\delta \sigma=0$, the relation~(\ref{eq:c12})
leads to (\ref{eq:delta-p-na-gsf}), as required.


\section{Condition for vanishing non-adiabatic pressure
perturbation}\label{sec:condition}

In Fourier space, $\nabla^{2}\Phi\propto \l(k^{2}\;\Phi_{k}\r)$.
It is then clear from Eq.~(\ref{eq:delta-p-na-gsf}) that the non-adiabatic
pressure perturbation $\delta p^{_{\rm NA}}$ will vanish in the super
Hubble limit (i.e. as $k\to 0$) for all scalar
fields~\cite{bep-shs,christopherson-2009}.
This occurs, for instance, when the modes are well outside the Hubble radius
during the inflationary epoch~\cite{bep-shs}.
It is interesting to enquire whether the non-adiabatic pressure perturbation
$\delta p^{_{\rm NA}}$ vanishes {\it identically}\/ for any scalar field model.
In such a case, the scalar field will behave {\it exactly}\/ like a perfect
fluid described by an equation of state, at least at the first order in the
perturbations.
It is evident from Eq.~(\ref{eq:delta-p-na-gsf}) that such a behavior will be
possible if and only if $c_{_{\rm S}}^2 = c_{_{\rm A}}^{2}$.
Based on this condition, we shall now arrive at the corresponding condition on
the Lagrangian density describing the scalar field.

Using Eq.~(\ref{eq:ce}) that describes the conservation of energy, and
the expressions~(\ref{eq:rho-p-gsf}) for the background energy density
and pressure associated with the scalar field, the adiabatic speed of
the perturbations~$c_{_{\rm A}}^{2}$ can be expressed as
\beq
c_{_{\rm A}}^{2}
=\l(\f{\dot p}{\dot \rho}\r)
=-\l(\f{\dot p}{\l(3\, H\r)\, \l(\rho+p\r)}\r)
= -\l(\f{{\ddot \phi}\,\l(\pa {\cal L}/\pa X\r)
+ \l(\pa {\cal L}/\pa \phi\r)}{\l(3\, H\, {\dot \phi}\r)\,
\l(\pa {\cal L}/\pa X\r)}\r).
\eeq
The equation of motion~(\ref{eq:em-gsf}) governing the field, then allows
us to arrive at the following expression for the adiabatic speed of the
perturbations:
\bea
c_{_{\rm A}}^{2}
&=&-\l(\l(3\, H\, {\dot \phi}\r)\, \l(\f{\pa {\cal L}}{\pa X}\r)\;
\l[\l(\f{\pa {\cal L}}{\pa X}\r)
+  (2\, X)\, \l(\f{\pa^{2} {\cal L}}{\pa X^{2}}\r)\r]\r)^{-1}\nn\\
& &\;\;\times\;
\biggl[2\, \l(\f{\pa {\cal L}}{\pa X}\r)\,
\l(\f{\pa {\cal L}}{\pa \phi}\r)
-(3\, H\, {\dot \phi})\, \l(\f{\pa {\cal L}}{\pa X}\r)^{2}
+ (2\, X)\, \l(\f{\pa^{2} {\cal L}}{\pa X^{2}}\r)\,
\l(\f{\pa {\cal L}}{\pa \phi}\r)
-(2\, X)\, \l(\f{\pa {\cal L}}{\pa X}\r)\,
\l(\f{\pa^{2} {\cal L}}{\pa X\;\pa \phi}\r)\biggr].\label{eq:ca2}
\eea
From the definition~(\ref{eq:cs2}) of $c_{_{\rm S}}^{2}$, it is
then straightforward to show that the condition $c_{_{\rm A}}^{2}
= c_{_{\rm S}}^{2}$ implies that
\beq
\l(\f{\pa {\cal L}}{\pa X}\r)\, \l(\f{\pa {\cal L}}{\pa \phi}\r)
+ X\; \l(\f{\pa^{2} {\cal L}}{\pa X^{2}}\r)\,
\l(\f{\pa {\cal L}}{\pa \phi}\r)
-X\; \l(\f{\pa {\cal L}}{\pa X}\r)\;
\l(\f{\pa^{2} {\cal L}}{\pa X\;\pa \phi}\r)= 0.\label{eq:condition}
\eeq
This condition can be further simplified to be
\beq
\f{\pa}{\pa X}\l[\l(\f{1}{Y}\r)\;
\l(\f{\pa {\cal L}}{\pa \phi}\r)\right]= 0,\label{eq:condition-sv}
\eeq
where we have defined $Y$ as
\beq
Y = X\;\l(\f{\pa {\cal L}}{\pa X}\r).
\eeq
As we had mentioned before, based on the analogy with fluids, we shall
refer to scalar fields that satisfy the condition~(\ref{eq:condition-sv})
as perfect scalar fields.
Interestingly, the condition~(\ref{eq:condition}) above has been arrived
at earlier while considering a completely different problem involving the
stationary configurations of scalar fields~\cite{akhoury-2009}}.
This seems to indicate some relation between stationarity and a vanishing
non-adiabatic pressure perturbation, which invites further investigation.

It is straightforward to show that, if a given Lagrangian density, say,
${\cal L}_{1}(X,\phi)$, satisfies the  condition~(\ref{eq:condition-sv}),
then the Lagrangian densities, say, ${\cal L}_{2}(X,\phi)$ and
${\cal L}_{3}(X,\phi)$, that are related to ${\cal L}_{1}(X,\phi)$ as
follows:
\beq
{\cal L}_{2}(X,\phi)
= C_{1}\; \exp\, \l[C_{2}\, {\cal L}_{1}(X,\phi)\r]
\qquad{\rm and}\qquad
{\cal L}_{3}(X,\phi)
= C_{3}\; {\rm log}\,\l[C_{4}\, {\cal L}_{1}(X,\phi)\r],
\label{eq:rbl}
\eeq
where $C_{1}$, $C_{2}$, $C_{3}$ and $C_{4}$ are constants, also satisfy
the same condition.
Though these three Lagrangian densities satisfy the
condition~(\ref{eq:condition-sv}), it may be worthwhile to note that,
they, in fact, lead to different evolution equations for the background
as well as the perturbations.


\section{Examples of perfect scalar fields}\label{sec:examples}

In this section, we shall construct examples of perfect scalar fields
that satisfy the condition~(\ref{eq:condition-sv}).
We shall first construct purely kinetic models that do not depend on the
scalar field directly, and then consider models that also seemingly depend
on the scalar field.


\subsection{Pure kinetic models}

Note that any Lagrangian density~${\cal L}$ that does not explicitly depend
on the scalar field~$\phi$, i.e. when $\l(\pa {\cal L}/\pa \phi\r) = 0$,
naturally satisfies the condition~(\ref{eq:condition-sv}).
These Lagrangian densities are often referred to as pure kinetic models.
It is well known that pure kinetic models can be described as barotropic,
perfect fluids which, by definition, do not possess any non-adiabatic
pressure perturbation (in this context, see
Refs.~\cite{tejedor-2005,bertacca-2008}).
Such models were originally considered in the context of inflation~\cite{ki},
and they have been resurrected more recently as a potential candidate of
dark energy~\cite{de-reviews}.
We shall now discuss a couple of specific examples of these models.


\subsubsection{A scalar field mimicking a perfect fluid with a constant
equation of state}

Consider a Lagrangian density of the following form:
\beq
{\cal L}(X) = V_{0}\; X^{\alpha},\label{eq:pf}
\eeq
where $V_{0}$ is a constant potential, and $\alpha$ is a real number.
Such a Lagrangian density clearly satisfies the
condition~(\ref{eq:condition-sv}).
The background energy density and pressure corresponding to this Lagrangian
density can be obtained to be [cf. Eq.~(\ref{eq:rho-p-gsf})]
\beq
\rho = \l(2\, \alpha - 1\r)\, V_{0}\; X^{\alpha}\qquad{\rm and}\qquad
p = V_{0}\; X^{\alpha},
\eeq
so that the resulting equation of state $w \equiv (p/\rho)$ is a constant,
and is given by
\beq
w = \l(\frac{1}{2\, \alpha - 1}\r).
\eeq
In other words, the scalar field described by the Lagrangian
density~(\ref{eq:pf}) essentially behaves like a perfect fluid with a constant
equation of state.
It is also useful to note that, in such a case, the two speeds of sound
simplify to [cf. Eqs.~(\ref{eq:cs2}) and (\ref{eq:ca2})]: $c_{_{\rm A}}^{2}
= c_{_{\rm S}}^{2} = w$.


\subsubsection{Models that behave as the Chaplygin gas}

Consider a tachyon that is governed by the Lagrangian density
\beq
{\cal L}(X) = -\l(V_{0}\; \sqrt{1 - 2\, X}\r),\label{eq:tm-cg}
\eeq
where $V_{0}$ is a constant potential.
Needless to add, this Lagrangian density also satisfies the
condition~(\ref{eq:condition-sv}).
The background energy density and pressure corresponding to the above
Lagrangian density are given by
\beq
\rho = \l(\f{V_{0}}{\sqrt{1 - 2\, X}}\r)\qquad{\rm and}\qquad
p = -\l(V_{0}\, \sqrt{1 - 2X}\r),
\eeq
so that we can write
\beq
p = -\l(\f{V_{0}^{2}}{\rho}\r).
\eeq
This is the equation of state that describes the Chaplygin
gas~\cite{kamenshchik-2001,bento-2002,chimento-2004,scherrer-2004}
and, in such a case, we obtain that
\beq
c_{_{\rm A}}^{2} = c_{_{\rm S}}^{2} = \l(1-2\, X\r)=-w.
\eeq
Upon using the conservation equation~(\ref{eq:ce}) for the background
energy density, it can be readily shown that the equation of state
parameter of the Chaplygin gas evolves as a function of the scale
factor in the following fashion:
\beq
w(a)
= -\l(\frac{V_{0}^{2}}{V_{0}^{2}+({\cal A}/a^{6})}\r),
\eeq
where ${\cal A}$ is a positive constant.

A more generic form of the Lagrangian density~(\ref{eq:tm-cg}) that
satisfies the condition~(\ref{eq:condition-sv}) is given by
\beq
{\cal L}(X)
= -\l(V_{0}\; \l[1 - \l(X/V_{0}\r)^{[(1
+ \alpha)/2\,\alpha]}\r]^{[\alpha/(1 + \alpha)]}\r),\label{eq:tm-gcg}
\eeq
where $V_{0}$ is again a constant, while $\alpha$ is a real number.
It is straightforward to show that, for this Lagrangian density, the
homogeneous energy density and pressure are given by
\beq
\rho=\l(V_{0}\; \l[1 - \l(X/V_{0}\r)^{[(1+ \alpha)/2\,
\alpha]}\r]^{-[1/(1 + \alpha)]}\r)
\qquad{\rm and}\qquad
p=-\l(V_{0}\; \l[1 - \l(X/V_{0}\r)^{[(1
+ \alpha)/2\,\alpha]}\r]^{[\alpha/(1 + \alpha)]}\r),
\eeq
so that, we have
\beq
p = -\l(\frac{V_{0}^{(1 + \alpha)}}{\rho^{\alpha}}\r),
\eeq
which is the equation of state of the generalized Chaplygin gas.
Over the last few years, such models have been considered in the literature
as possible candidates for dark
energy~\cite{bento-2002,chimento-2004,scherrer-2004}.
In this case, the equation of state parameter $w$ evolves as
\beq
w(a)
= -\l(\frac{V_{0}^{(1 + \alpha)}}{V_{0}^{(1 + \alpha)}
+\l({\cal A}/a^{[3\, \l(1+\alpha\r)]}\r)}\r),
\eeq
where ${\cal A}$ is again a positive constant.
Also, we find that
\beq
c_{_{\rm A}}^{2} = c_{_{\rm S}}^{2}
= \alpha\; \l[1 - \l(X/V_{0}\r)^{[(1+ \alpha)/2\,\alpha]}\r].
\eeq


\subsubsection{Another model}

Now, consider a Lagrangian density of the following form:
\beq
{\cal L}(X)
= -V_{0}\l(\sqrt{X}- 1\r)^{-\alpha},\label{eq:dm-de}
\eeq
where, again, $V_{0}$ is a constant, and $\alpha$ is a real number.
The background energy density and pressure corresponding to the
above Lagrangian density are found to be
\beq
\rho = V_{0}\, \l[(1+\alpha)\, \sqrt{X} -1\r]\;
\l(\sqrt{X}- 1\r)^{-(1+\alpha)}
\qquad{\rm and}\qquad
p = -V_{0}\, \l(\sqrt{X}- 1\r)^{-\alpha}.\label{eq:dm-de-rho-p}
\eeq
In such a case, it can be shown that the equation of state parameter
$w$ evolves with the scale factor as
\beq
w(a)=-\l(\f{1}{\l(1+\alpha\r)+\l(\alpha/{\cal A}\r)\,
a^{-\l[3/(1+\alpha)\r]}}\r),\label{eq:w-another-model}
\eeq
where, as before, ${\cal A}$ is a positive constant, while
\beq
c_{_{\rm A}}^{2} = c_{_{\rm S}}^{2}
=\l(\f{1-\sqrt{X}}{\l(1+\alpha\r)\, \sqrt{X}}\r).
\eeq
Note that the expression~(\ref{eq:w-another-model}) for $w(a)$ implies that,
for positive $\alpha$, $w(a)$ evolves from zero to an asymptotic value of
$-(1 + \alpha)^{-1}$.
For an appropriate choice of the constants ${\cal A}$ and $\alpha$, it is then
possible to achieve $w(a)$ at the present epoch to be less than $-(1/3)$.
Hence, the above Lagrangian density may be considered as a possible model of dark
energy.


\subsection{Masquerading scalar fields}

We find that we can construct three types of models that {\it seemingly}\/
depend on the scalar field and also satisfy the
condition~(\ref{eq:condition-sv}).

The first type are models wherein the Lagrangian density of the scalar
field can be expressed as a sum of the kinetic and the potential terms
as follows:
\beq
{\cal L}(X,\phi) = f(X)-V(\phi).\label{eq:one}
\eeq
In such a situation, the condition~(\ref{eq:condition-sv}) immediately
restricts the function $f(X)$ to be
\beq
f(X)=\alpha\; {\rm log}\, \l(X/X_{0}\r),\label{eq:two}
\eeq
where $\alpha$ and $X_{0}$ are positive constants.
Moreover, the energy density and pressure associated with the homogeneous
scalar field are given by
\beq
\rho=\,\alpha\, \l[2-{\rm log}\, \l(X/X_{0}\r)\r]+V(\phi)
\qquad{\rm and}\qquad
p =\alpha\; {\rm log}\, \l(X/X_{0}\r)-V(\phi),
\eeq
and we find that the equation of state parameter evolves as
\beq
w(a) = -1  +\l(\frac{2\, \alpha}{{\cal A} - 6\, \alpha\, \log\, a}\r),
\eeq
where ${\cal A}$ is a suitably chosen positive constant.
Also, we obtain that $c_{_{\rm A}}^{2} = c_{_{\rm S}}^{2} =-1$.

The second type of models are wherein we can write the Lagrangian
density as a product of the kinetic and the potential terms in the
following fashion~\cite{ki,chimento-2004,scherrer-2004}:
\beq
{\cal L}(X,\phi) = f(X)\, V(\phi).\label{eq:ttm}
\eeq
In such cases, the condition~(\ref{eq:condition-sv}) restricts
the function~$f(X)$ to be $X^{\alpha}$, where $\alpha$ is a real
number.
The smooth energy density and pressure associated with this Lagrangian
density can be obtained to be
\beq
\rho=(2\, \alpha-1)\, \l[X^{\alpha}\, V(\phi)\, \r]
\qquad{\rm and}\qquad p=\l[X^{\alpha}\,V(\phi)\, \r].
\eeq
The resulting equation of state is evidently a constant, and is
given by
\beq
w = \l(\frac{1}{2\, \alpha - 1}\r).
\eeq
Obviously, this example is a more general case of the Lagrangian
density~(\ref{eq:pf}) we had discussed before.
As in the earlier example, the adiabatic and the effective speeds of
sound associated with the perturbations reduce to
\beq
c_{_{\rm A}}^{2} = c_{_{\rm S}}^{2} = \l(\f{1}{2\,\alpha-1}\r)=w.
\eeq
Note that the two Lagrangian densities that we have considered in
this sub-section until now can be related to each other through the
relations~(\ref{eq:rbl}).

The third type of scalar fields that we shall consider are those
that are described by the Lagrangian density
\beq
{\cal L}(X,\phi) = {\cal B}\, \exp\, \l[X^{\alpha}\, V(\phi)\r],
\label{eq:gm}
\eeq
where ${\cal B}$ is a negative constant, and $\alpha$ is a real
number.
Clearly, as the previous example, this Lagrangian density too
will satisfy the condition~(\ref{eq:condition-sv}), courtesy of
the relations~(\ref{eq:rbl}).
We obtain the homogeneous energy density and pressure associated
with the above Lagrangian density to be
\beq
\rho= {\cal B}\,\, \l[2\,\alpha\, X^{\alpha}\, V(\phi)-1\r]\;
\exp\, \l[X^{\alpha}\, V(\phi)\r]\,
\qquad{\rm and}\qquad
p={\cal B}\, \exp\, \l[X^{\alpha}\,V(\phi)\,\r].
\eeq
Also, we find that
\beq
c_{_{\rm A}}^{2} = c_{_{\rm S}}^{2}
= \l(\f{1}{2\,\alpha\, \l(1+V(\phi)\, X^{\alpha}\r)-1}\r),
\eeq
so that we can write
\beq
c_{_{\rm S}}^{2} = \l(\frac{w}{2\,\alpha\, w + 1}\r).
\eeq
Note that, if $\alpha > (3/2)$, then $c_{_{\rm S}}^{2}$ is a positive
definite quantity, and $w$ can be less than $-(1/3)$.

Though the above three examples behave as though they explicitly depend on
the scalar field, we find that they can be turned into pure kinetic models
by a suitable redefinition of the field.
For instance, the transformation
\beq
\chi(\phi)=\int\; d\phi\; \l[\f{V(\phi)}{V_{0}}\r]^{(1/2\,\alpha)},
\eeq
reduces the Lagrangian density~(\ref{eq:ttm}) to the example~(\ref{eq:pf}).
Evidently, such a transformation will also convert the
Lagrangian density~(\ref{eq:gm}) into a purely kinetic form.
Moreover, we find that the following transformation
\beq
\chi(\phi)=\int\; d\phi\; \exp-\l[V(\phi)/2\,\alpha\r],
\eeq
removes the explicit dependence on the scalar field in the first example
above [cf.~Eqs.~(\ref{eq:one}) and (\ref{eq:two})].
In other words, classically, the three Lagrangian densities that we have
presented in this sub-section are essentially pure kinetic models that
masquerade as though they depend on the scalar field.
The fact that field redefinitions can relate different, but dynamically
equivalent, Lagrangian densities for scalar fields has been noticed
earlier (in this context, see, in particular, Ref.~\cite{bertacca-2008}).
However, we should add that it is not clear whether such a behavior will
be preserved when the quantum nature of the scalar fields are taken into
account, as it is required, say, in the context of inflation.

In this section, we had presented a few specific examples of scalar fields
that satisfy the condition~(\ref{eq:condition-sv}).
It should be noted that the most general solution to this equation can be
expressed in terms of the product $\l[X\;f(\phi)\r]$, where $f(\phi)$ is
an arbitrary function of the scalar field (for details, see
Ref.~\cite{akhoury-2009}).


\section{An interesting implication for inflation}\label{sec:ii}

We find that the perfect scalar fields that satisfy the
condition~(\ref{eq:condition-sv}) behave in a particular fashion
when they are treated as the inflaton.
We shall now turn to discuss this property.

It is well known that, in the slow roll inflationary scenario
involving a single scalar field, the amplitude of the curvature
perturbation approaches a constant value soon after the modes
leave the Hubble radius (see, any of the standard
texts~\cite{texts} or reviews~\cite{reviews}).
However, this is not true when there are one or more brief periods of
deviation from slow roll inflation\footnote{Actually, it could even
be a short period of departure from inflation.
But, the deviation from slow roll {\it has}\/ to be brief, since a
prolonged deviation may not allow inflation to restart.}.
In such cases, the amplitude of the curvature perturbation over a
suitable range of modes (which leave the Hubble radius just before
the deviation from slow roll) can be enhanced or suppressed at
super Hubble scales when compared to their value at Hubble exit.
It can be shown that this behavior essentially arises due to the fact
that, in such scenarios, the intrinsic entropy perturbation (i.e. the
non-adiabatic pressure perturbation) grows rapidly around the period of
fast roll at super Hubble scales~\cite{bep-shs}.

Since the non-adiabatic pressure perturbation vanishes identically,
the phenomenon described above can not occur in models which satisfy
the condition~(\ref{eq:condition-sv}).
It is then possible that the demand that the non-adiabatic pressure
perturbation vanishes exactly is so restrictive that the perfect
scalar fields will not allow a brief period of departure from slow
roll inflation.
This expectation indeed turns out to be true and, in what follows, we
shall outline its proof.

Recall that the first slow roll parameter~$\epsilon$ is defined
as~\cite{texts,reviews}
\beq
\epsilon = -\l(\f{\dot H}{H^2}\r)=\l(\f{3\, (1 + w)}{2}\r),
\eeq
and inflation corresponds to the epoch wherein $\epsilon<1$ or,
equivalently, $w< -(1/3)$.
The time derivative of the quantity~$\epsilon$ is given by
\beq
{\dot \epsilon}
= \l(\f{9\, H}{2}\r)\, \l(1+w\r)\, \l(w-c_{_{\rm A}}^2\r)
=\l(\f{9\, H}{2}\r)\, \l(1+w\r)\, \l(w-c_{_{\rm S}}^2\r),
\label{eq:epsilon-dot}
\eeq
where, in arriving at the last equality, we have set $c_{_{\rm A}}^2
= c_{_{\rm S}}^{2}$, as it is in the case of perfect scalar fields.
The scalar power spectrum is determined by the amplitude of the curvature
perturbations when the modes are well outside the Hubble radius during the
inflationary epoch~\cite{texts,reviews}.
Also, these curvature perturbations are evolved from the sub Hubble to
the super Hubble scales by imposing standard initial conditions (viz.
the Bunch-Davies conditions) when the modes are well inside the Hubble
radius.
It is the quantity $c_{_{\rm S}}$ that describes the speed of propagation
of the curvature perturbations.
Hence, $c_{_{\rm S}}$ has to be a real quantity, if it has to be ensured
that the curvature perturbations do not grow exponentially at sub Hubble
scales, so that the Bunch-Davies initial conditions can be imposed on
them~ (see Ref.~\cite{ki}; in this context, also see Ref.~\cite{shanki-2009}).
So, during inflation, in addition to $w<-(1/3)$, we require that
$c_{_{\rm S}}^{2}>0$.
It is then evident from the last equality in the
expression~(\ref{eq:epsilon-dot}) above that
\beq
{\dot \epsilon}<0.
\eeq
In other words, $\epsilon$ will {\it always}\/ be a monotonically
{\it decreasing}\/ function of time when perfect scalar fields drive
inflation.
Therefore, clearly, once slow roll inflation has commenced, these models
will not allow a period of deviation from slow roll inflation.
In fact, terminating inflation becomes a problem in such cases, and
one will have to invoke an additional scalar field to exit inflation.
We should point out that, amongst the various examples of perfect
scalar fields that we have discussed, only the models described in Eqs.\~(\ref{eq:tm-cg}),
(\ref{eq:tm-gcg}) and~(\ref{eq:gm}) satisfy both the conditions $w<-(1/3)$
and $c_{_{\rm S}}^{2}>0$ to act as a inflaton.

It has been recognized that certain features in the
primordial spectrum lead to a better fit to the cosmic microwave
background data than the conventional, featureless, power law,
primordial scalar perturbation spectrum (in this context, see, for
example, Refs.~\cite{fips}, and the long list of references therein).
Interestingly, a short epoch of deviation from slow roll inflation
is essential for generating features in the inflationary, scalar
perturbation spectrum.
In particular, it seems mandatory that the first slow roll parameter
$\epsilon$ has to rise and fall if features are to be produced in the
scalar power spectrum\footnote{For example, it is known that the
spectrum can remain featureless even if the second slow roll
parameter is large for a short duration of time (see, for
instance, Refs.~\cite{kinney}).}.
However, as we discussed above, it is not possible to achieve such
a behavior for $\epsilon$ in perfect scalar field models.
This, in turn, implies that scalar fields with a vanishing non-adiabatic
pressure perturbation, if they are treated as the inflaton, can not lead
to features in the scalar perturbation spectrum!


\section{Concluding remarks}\label{sec:cr}

In this paper, we have obtained a condition on the Lagrangian density
of an arbitrary scalar field under which the intrinsic, non-adiabatic
pressure perturbation associated with the scalar field vanishes
identically, at the first order in the perturbations.
Motivated by the analogy with fluids, we have termed these fields as
perfect scalar fields.
Scalar field models that depend only on the kinetic energy are known
to contain no non-adiabatic pressure perturbation.
In addition to discussing a couple of such examples, we have presented
a few models which seemingly depend on the scalar field explicitly, and
possess no non-adiabatic pressure perturbation.
But, we find that all such models that we were able to construct could
be reduced to purely kinetic models by a redefinition of the field.
In fact, it can be proved that all Lagrangian densities that satisfy the
condition~(\ref{eq:condition-sv}) can be reduced to a purely kinetic form
(in this context, see the discussion following Eq.~(2.22) in
Ref.~\cite{akhoury-2009}).
Interestingly, we have also shown that, if perfect scalar fields are
used to drive inflation, then they can not lead to features in the scalar
power spectrum.

Note that, throughout the paper, we had restricted our attention to the
first order in the perturbations.
We should point out that, since all the scalar fields that satisfy the
condition~(\ref{eq:condition-sv}) can be reduced to pure kinetic models,
such perfect scalar fields will not contain any non-adiabatic pressure
perturbation at the higher orders in the perturbations either (in this
context, see Refs.~\cite{sopt}).
In other words, these scalar fields are truly perfect.


\section*{Acknowledgments}

LS and SU wish to thank the Inter University Centre for Astronomy and
Astrophysics, Pune, India, and the Harish-Chandra Research Institute,
Allahabad, India, for hospitality, where this work was initiated and
completed, respectively.
LS also wishes to thank Will Kinney and Jerome Martin for discussions,
and Hermano Velten for drawing attention to a few relevant references.


\end{document}